\documentclass[a4paper]{elsarticle}

\usepackage{color}
\usepackage{xspace}

\usepackage{inputenc}
\usepackage{amsmath}
\usepackage{amsfonts}
\usepackage{amssymb}
\usepackage{enumerate}
\usepackage{natbib}
\usepackage{ifdraft}
\usepackage{color}
\usepackage{epsfig}
\usepackage[dvips]{geometry}
\usepackage{pdfpages}
\usepackage{pst-math}
\usepackage{txfonts}
\usepackage{times}
\usepackage{times}

\makeatletter
\def\@author#1{\g@addto@macro\elsauthors{\normalsize%
    \def\baselinestretch{1}%
    \upshape\authorsep#1\unskip\textsuperscript{%
      \ifx\@fnmark\@empty\else\unskip\sep\@fnmark\let\sep=,\fi
      \ifx\@corref\@empty\else\unskip\sep\@corref\let\sep=,\fi
      }%
    \def\authorsep{\unskip,\space}%
    \global\let\@fnmark\@empty
    \global\let\@corref\@empty  
    \global\let\sep\@empty}%
    \@eadauthor={#1}
}
\makeatother

\renewcommand{\bibname}{References}%

\let\ElseVierBibliography\bibliography%
\renewcommand{\bibliography}[1]{%
\section*{\bibname}%
\ElseVierBibliography{#1}%
}

\newtheorem{thm}{Theorem}
\newtheorem{proof}{Proof}
\newtheorem{lem}{Lemma}
\newtheorem{prop}{Proposition}

\newtheorem{coll}{Corollary}

\newcommand{\Banach}{\mathfrak{B}}
\newcommand{\Hilbert}{\mathcal{H}}

\newcommand{\Real}{\mathbb{R}}
\newcommand{\complex}{\mathbb{C}}
\newcommand{\Mat}{\mathsf{M}}
\newcommand{\Her}{\mathsf{H}}
\newcommand{\op}[1]{\mathbf{#1}}
\newcommand{\Tr}[1]{\text{Tr}\left( #1\right)}
\newcommand{\HS}[2]{\left(#1, #2\right)_{HS}}
\newcommand{\normHS}[1]{\left\Vert #1\right\Vert_{HS}}
\newcommand{\iden}{\mathbf{I}}
\newcommand{\X}{\vec{x}}
\newcommand{\f}{\vec{f}}
\newcommand{\ENORM}[1]{\left\vert #1 \right\vert}
\newcommand{\norm}[1]{\left\Vert #1 \right\Vert}
\newcommand{\proj}{\mathbf{P}}

\newcommand{\totmixed}{\frac{1}{d}\iden_d}

\newcommand{\st}{\mathsf{S}}
\newcommand{\T}{\op{M}}

\newcommand{\rot}{\mathbf{R}}
\newcommand{\U}{\mathcal{U}}

\newcommand{\SO}{{SO}}
\newcommand{\ls}{S_L}
\newcommand{\vs}{S_{vN}}
\newcommand{\pos}{\mathcal{P}}
\newcommand{\scale}{\mathbf{S}}
\newcommand{\V}{\op{R}}
\newcommand{\pp}{\op{S}}

\newcommand{\Lind}{\mathcal{L}}
\newcommand{\diss}{\mathcal{D}}

\newcommand{\M}{\op{M}}

\newcommand{\matL}{\mathbf{L}}
\newcommand{\genL}{\Lambda}

\newcommand{\K}{\op{K}}

\newcommand{\Hamil}{\mathbf{H}}

\biboptions{sort&compress}

\begin{document}

\begin{frontmatter}



\title{Unitary-Scaling Decomposition and Dissipative Behaviour in Finite-Dimensional Unital Lindblad Dynamics}

\author{Fattah Sakuldee\corref{cor1}}
\ead{fattah.sak@student.mahidol.ac.th}
\cortext[cor1]{Corresponding author}

\author{Sujin Suwanna}
\ead{sujin.suw@mahidol.ac.th}

\address{MU-NECTEC Collaborative Research Unit on Quantum Information, Department of Physics, Faculty of Science, Mahidol University, Bangkok, 10400, Thailand.}


\begin{abstract}
We investigate a decomposition of a unital Lindblad dynamical map of an open quantum system into two distinct types of mapping on the Hilbert-Schmidt space of quantum states. One component of the decomposed map corresponds to reversible behaviours, while the other to irreversible characteristics. For a finite dimensional system, we employ real vectors or Bloch representations and express a dynamical map on the state space as a real matrix acting on the representation. It is found that rotation and scaling transformations on the real vector space, obtained from the real-polar decomposition, form building blocks for the dynamical map. Consequently, the change of the linear entropy or purity, which indicates dissipative behaviours, depends only on the scaling part of the dynamical matrix. The rate of change of the entropy depends on the structure of the scaling part of the dynamical matrix, such as eigensubspace partitioning, and its relationship with the initial state. In particular, the linear entropy is expressed as a weighted sum of the exponential-decay functions in each scaling component, where the weight is equal to $\vert\vec{x}_k(\rho)\vert^2$ of the initial state $\rho$ in the subspace. The dissipative behaviours and the partition of eigensubspaces in the decomposition are discussed and illustrated for qubit systems.
\end{abstract}


\end{frontmatter}



\section{Introduction}
A dynamical map to describe the evolution of a system between two given time epochs is one of the most versatile mathematical objects used in physics, especially in quantum physics \cite{Neumann,Alicki}. In a close quantum system, such a dynamical map is described by a strongly continuous one-parameter unitary group on the operator space or on its dual as inspired by the celebrated work of von Neumann \cite{Neumann}. Despite the fact that this formulation sets the fundamental framework for dynamical analysis in quantum physics, it has been shown to be limited by many restrictions. For instances, the system may not inherit the closeness property; or it may not reach an equilibrium state in finite time; or the time homogeneity of the dynamics may not hold \cite{DeffnerLutz2011,Leifer2013,grech,Gogolin2011,Baiesi2010}. Consequently, many extensions have been proposed to model quantum systems such as open systems, or those in non-equilibrium or non-stationary regimes; see \cite{2013arXiv1310.1484F,Baumgartner2014,Pucci,Baiesi2013,Kavan2012,Alicki,PhysRev.121.920} for more details on the development of this subject. 

It is commonly known that an open quantum system is not governed by the \\Schr\"odinger-type differential equation since the unitary evolution arising from such the equation cannot adequately explain irreversible behaviours of the dynamics. However, when the Markovian property is assumed (i.e. the dynamics are time homogeneous; see Section \ref{subsec:markov}), the alternative formulations have been successfully achieved to sufficiently explain physical phenomena in open quantum systems. These include the Kossakowski-Gorini-Sudarshan formulation of the dynamical maps for finite dimensional open quantum systems \cite{GKS}; the more-general Lindblad formulation which includes the dynamics on infinite dimensional systems \cite{Lindblad}; and the Davies' or Nakajima-Zwanzig constructions which interpret the Markovian property as a result of a certain limit in the composite dynamics \cite{davies1974,Rivas,Alicki}.

For open quantum systems, any dynamical map can be characterised into three categories: (i) the Liouville-von Neumann type, where the dynamics are precisely described by unitary groups \cite{Neumann}; (ii) the Lindblad type, where the complete positivity of a map and the Markovian property are required, while decay, decoherence or dephasing are allowed \cite{Lindblad,Baumgartner2008}, and the dynamical maps can be treated as an ultra-weakly continuous one-parameter semigroup on the operator space; and (iii) the beyond-Lindblad type, where the assumptions on complete positivity or the Markovian property are relaxed \cite{PhysRevA.77.042113,Alicki}. Among these types of the dynamical maps, we observe that the most significant different characteristics is the entropy change with the dynamics \cite{Wehrl1978,geobook}. In the Liouville-von Neumann type, the entropy change is zero by the unitary invariance, whereas in the Lindblad type, the entropy change essentially increases and possesses an asymptote, signalling a steady state in thermalization and relaxation of the system \cite{Baumgartner2008,Pucci,Diertz}. 

Interestingly, the entropy change beyond the Lindblad dynamics remains open. This leads us to investigate the entropy change as the characterisation parameter of the quantum dynamics. Our interests lie in the Lindblad dynamics and beyond of open quantum systems. We hypothesise that the quantum dynamics should be characterised by two parameters, one for a reversible or coherence process, and the other for an irreversible or decoherence process. While this idea follows from intuition, it has not been explicitly verified or used to extract information about the evolution of an open quantum system. In particular, the relationship between the reversible and irreversible components of the process has never been explicitly derived on the level of a dynamical map, rather than that of a generator. This technique can be beneficial in case that the dynamical map does not obviously exhibit a generator, for example, a dynamical map derived from a process tomography.

A key idea of this work lies in a well-known polar decomposition of a matrix, which allows us to perform the unitary-scaling decomposition of a dynamical map. In a nutshell, we characterise the concerning dynamical matrices into two types: the rotation matrix describing the unitary or coherence behaviour, and the scaling matrix describing the dissipative or decoherence behaviour of the dynamics. As a general goal, we conjecture that such a decomposition should be valid in all three mentioned types of dynamics. However, for various technical difficulties (see Section \ref{sec:scope} for more discussion), we have presented here the results for the unital Lindblad dynamics in a finite-dimensional open quantum system, where we obtain an exact relation between rotation and scaling components, as well as the effects of the initial state in the entropy change. In essence, the entropy change depends on not only the dynamical map and the initial state, but also the interplay relationship between them, i.e. the linear entropy will change with different rates if the initial state is prepare in different subspaces of the dynamical map. We show that this assertion is valid generally for a normal dynamical map.

The article is organised into sections as follow. Section \ref{sec:representation} contains necessary background and relevant mathematical preliminaries of a dynamical map of an open quantum system. This section also includes the scope and the discussion of important assumptions for our current work. Based the polar decomposition of a matrix, the unitary-scaling scaling decomposition of a dynamical map is derived in Section \ref{sec:decom}. In Sections \ref{sec:observe} and \ref{sec:entropy}, the consequent results are presented, where in Section \ref{sec:observe}, we focus on the contribution to the dynamics in the isotropic scaling case. More importantly, the results on entropy change and characteristic of dissipative behaviours are presented in Section \ref{sec:entropy}. The application of the decomposition to qubit systems are remarked in both Sections \ref{sec:observe} and \ref{sec:entropy}. We note that we illustrate the dissipative behaviour of the dynamics by using the linear entropy, which retrieves the characteristics of the Lindblad dynamics as expected. Finally the conclusions are summarised in Section \ref{sec:discussion}. Additionally, relevant information and results on simple examples of elementary physical processes in qubit systems can be found in Appendices.

\section{Mathematical Preliminaries}\label{sec:representation}

\subsection{Matrix and General Bloch Representation}
Finite-dimensional quantum systems can be commonly represented by real vectors (aka Bloch vector representation, or coherent representation) \cite{Kimura2003339,Kosakowski2003}. The state-observable description is given by a pair $(\rho,\op{a})$ of a density matrix $\rho\in\st_d:=\{\rho\in\Mat_d(\complex):\rho\geq 0,\Tr{\rho}=1\}$, and an observable
$\op{a}\in\Her_d := \{ \op{a}\in\Mat_d(\complex): \op{a}=\op{a}^* \}$ \cite{Lassner}, where $\Mat_d(\complex)$ denotes the set of $d\times d$ complex matrices equipped with the Hilbert-Schmidt inner product $\HS{\op{a}}{\op{b}} := \Tr{\op{a}^* \op{b}}$ and with norm $\normHS{\op{a}}:=\sqrt{\HS{\op{a}}{\op{a}}}.$ By choosing an orthogonal basis set $F=\{f_\alpha\}$ of $\Her_d$ with $f_0:=\frac{1}{\sqrt{d}}\iden_d$ and $\Tr{f_\alpha} := 0$, for $\alpha=1,\ldots,d^2-1,$ we can write a real vector representation for $\op{a}$ in $\Real^{d^2-1}$ as 
$\X(\op{a}):=\left( x_1(\op{a}), x_2(\op{a}), \ldots, x_{d^2-1}(\op{a})\right)\in\Real^{d^2-1},$ 
where $x_\alpha(\op{a}):=\HS{\op{a}}{f_\alpha}=\Tr{\op{a} f_\alpha}.$ We can therefore obtain 
	\begin{equation}
		\op{a} = \frac{\Tr{\op{a}}}{d}\iden_d + \f\cdot\X(\op{a}),
		\label{eq:BlochForm}
	\end{equation}
where $\f=\left(f_1,f_2,\ldots,f_{d^2-1} \right)$ is a $(d^2-1)$-tuplet of matrix bases $f_\alpha$. For a density operator $\rho,$ its real vector representation can be expressed as in Eq. \eqref{eq:BlochForm} with $\Tr{\rho} = 1.$ 

Let $\Phi$ denote a dynamical map. This linear map can be represented as a matrix acting on the generalised Bloch vector. We assume some physical conditions on $\Phi :$
	\begin{enumerate}[(C1)]
		\item $\Phi$ is completely positive and trace-preserving in the sense that $\Phi\otimes\iden_n$ maps a positive operator to a positive operator for all integers $n$ and $\Tr{\Phi(\rho)}=\Tr{\rho}$ for all $\rho\in\st_d.$ \label{as:PTP}
		\item $\Phi$ is contractive in the Hilbert Schmidt norm, i.e. $\normHS{\Phi(\rho)}\leq\normHS{\rho}$ for all $\rho\in\st_d.$ \label{as:contract}
	\end{enumerate}
The condition (C\ref{as:PTP}) is intuitive for a  dynamical map since the physical dynamics should preserve the complete positivity and normalisation of probability. 
In addition to complete positivity, for simplicity, the condition (C\ref{as:contract}) has been assumed in this work. Even condition (C\ref{as:contract}) is not a common nature of dynamical maps, it holds in an open quantum system undergoing unital dynamics, e.g. absence of translation. 
From the condition (C\ref{as:contract}), note that the unital property is coincided with contractivity in the Hilbert-Schmidt norm \cite{WangSchirmer2009}. See Section \ref{sec:scope} for more discussion. That the conditions (C\ref{as:PTP}) and (C\ref{as:contract}) are consistent with the properties of the Lindblad dynamics can be seen in \ref{appen:Lind}.

In finite dimensions, a unital dynamical map $\Phi$ can be expressed as
	\begin{equation}
		\Phi (\rho) := \varphi[\T](\rho) = \totmixed + \f\cdot\left( \T\cdot\X(\rho) \right), 
		\label{eq:counterpartmap}
	\end{equation}
where $\T$ is a linear transformation on $\Real^{d^2-1}$ \cite{Kosakowski2003}. From Eq. \eqref{eq:counterpartmap}, it follows that the positivity and contraction conditions of $\T$ in the Euclidean norm $\ENORM{\T\cdot\X}\leq\ENORM{\X}$ correspond to those of the corresponding map $\varphi[\T]$ by the assumptions (C\ref{as:PTP}) and (C\ref{as:contract}). Moreover, the trace preserving property of $\varphi[\T]$ follows from the traceless property of the bases $f_\alpha$. The dynamical map in Eq.\eqref{eq:counterpartmap} satisfies both conditions (C\ref{as:PTP})-(C\ref{as:contract}), and this equation allows us to consider the map on the real vector representation of the states instead. We also note that the mapping $\T\mapsto\varphi[\T]$ in Eq. \eqref{eq:counterpartmap} is multiplicative in the sense that $\varphi[\op{A}\op{B}]=\varphi[\op{A}]\circ\varphi[\op{B}].$

\subsection{Remarks on Unital and Normal Properties of Dynamical Maps}\label{sec:scope}

In our current work, we are strictly working with unital dynamical maps. We will briefly discuss our framework and assumptions. This unitality assumption concerns the dynamical equilibrium issue. Specifically, the dual map $\Phi^*$ of $\Phi$ is defined by $\Tr{\rho\Phi^*\left(\op{a}\right)} = \Tr{\Phi\left(\rho\right)\op{a}}$ for any operator $\op{a}$ and a density operator $\rho.$ If $\Phi$ is unital, it is equivalent to that $\Phi^*$ is trace preserving, i.e.  $\Tr{\Phi^*\left(\op{a}\right)} = \Tr{\op{a}}$. That implies that there is no \emph{effective} transfer of any physical quantity, including energy and the number of particles, between the system and the environment. In other words, all physical quantities are the integrals of motion.

On the other hand, for a non-unital dynamical map, there appears to have multiple possibilities to interpret the structure and nature of the dissipative behaviours. For example, one can introduce the translation matrix representing the image of a totally mixed state $\iden_d/d$. In this case, the action of $\T$ includes a translation part in Eq. \eqref{eq:counterpartmap}:
	\begin{equation}
		\Phi (\rho) := \varphi[\T](\rho) = \totmixed + \f\cdot\left( \T\cdot\X(\rho) + \vec{c} \right) = \totmixed + \f\cdot\vec{c} + \f\cdot\left( \T\cdot\X(\rho) \right), 
		\label{eq:translation}
	\end{equation}
where the translation vector $\vec{c}$ can be time dependent. The shift by $\f\cdot\vec{c}$ destroys the multiplicative property of the dynamical map. Please see \ref{appen:SR-TSR} for more details. 
Note that the representation still takes place on the Hilbert-Schmidt space, which is a compact subset of $\Real^{d^2-1}.$ 
Another example, one can avoid this difficulty in the first place and directly use the canonical bases inherited from the Hilbert space (aka matrix elements of $d^2$ dimensions on the complex field) and then analyse the dissipative behaviour from some measures, such as entropy production or fidelity \cite{PhysRevA.70.052309,Baumgartner2008,Baumgartner2008b,Sudarshan2003}. 
The advantages and disadvantages of both approaches are discussed in \ref{appen:SR-TSR}. For simplicity, we will only consider the unital dynamical maps in our current investigation. 

Another property assumed is the normality condition of the dynamical map. Obviously, this property can be expressed in terms of diagonalisability of the dynamical matrix, which leads to the structure of the subspaces corresponding to the eigenvectors of the dynamical matrix. In literature, the normality property also relates to the existence and accessibility of a steady state. For instance, it has been shown that the normality property is sufficient for the dynamical matrix to be contractive with respect to the Hilbert-Schmidt norm \cite{WangSchirmer2009}. For the governing equation of the Lindblad type, one can also observe that the condition for the existence and accessibility of steady state (Spohn's theorem) will also force the dynamical map to be normal as well \cite{Rivas}. Loosely speaking, it can be said that the normal property is sufficient for the dynamical map to reach its steady state. Whether it is also the necessity condition remains an open question, and is not part of the current research. 

Consequently, in this paper, we restrict ourselves to investigate only the unital and normal dynamical maps for simplicity and to avoid some technical difficulties mentioned above. We should point out that these two conditions are different conditions; namely, the normality property does not imply that of unitality, and vice versa.

\subsection{Geometric Characteristics of Translation, Scaling and Unitary Maps}

A unitary transformation $\U$ on the state space $\st_d$ can be represented by a rotation matrix $\rot$ on the vector space $\Real^{d^2-1}$ \cite{Alicki,geobook,NC} as 
	\begin{equation}
		\U(\rho) := \varphi[\rot](\rho) = \totmixed  + \f\cdot(\rot\cdot\X(\rho)).
	\label{eq:rotate}
	\end{equation}
This transformation preserves the traces of $\rho$ and $\rho^2$ since $\Tr{f_\alpha}=0$ for $\alpha=1,\ldots,d^2-1$, and $\rot$ preserves the Euclidean norm. The latter remains the same in a unitary process, indicating the unchanged degree of mixing of the state. This can be seen from the linear entropy $\ls(\rho):=1-\Tr{\rho^2}$, which is also a leading term of the von-Neumann entropy $\vs(\rho):=-\Tr{\rho\ln\rho}$. See more details in Section \ref{sec:entropy}.

A scaling map $\scale$ is another type of the map in an open quantum system. We define the diagonal scaling matrix $\scale_D$ by
	\begin{equation}
		\scale_D	\cdot\X	=	 \sum_{\alpha=1}^{d^2-1}e^{-\lambda_\alpha}\proj^G_\alpha(\X),
		\label{eq:scale_D}	
	\end{equation}
where $G=\{f_0\}\cup\{g_\alpha\}_{\alpha=1}^{d^2-1}$ is another orthogonal basis set of $\Her_d$ related to $F$ by the Jacobian matrix $\op{J};$ and $\proj^G_\alpha(\X):= x^G_\alpha g_\alpha.$ Here $\X^G$ is a vector in $\Real^{d^2-1}$ with $x^G_\alpha$ being a norm of $\X(\totmixed + g_\alpha);$ and $\lambda_\alpha\geq 0 $ for all $\alpha=1,\ldots,d^2-1$ since we consider only positive and contraction maps. Commonly, the basis set $G$ is known in literature as the damping bases \cite{BriegelEA1993,DariuszEA2010}. In the basis $F,$ the scaling map can be represented as
	\begin{equation}
		\scale	=	\op{J}^{-1}\scale_D\op{J}.
		\label{eq:scale}
	\end{equation}
Let $\pos$ denote a map on $\st_d$ induced by the scaling matrix $\scale.$  The map $\pos$ is called \emph{isotropic} if all scaling parameters $\lambda_\alpha$ are identical; otherwise, it is called \emph{anisotropic}. For the isotropic case, setting $\lambda_\alpha=\lambda$ for $\alpha=1,2,\ldots,d^2-1$, we obtain
	\begin{equation}
		\pos_\lambda(\rho) := e^{-\lambda}\rho+\left(1-e^{-\lambda}\right)\totmixed.
		\label{eq:isotropic_map}
	\end{equation}
All positive symmetric contraction matrices on $\Real^{d^2-1}$ have corresponding scaling transformations, since they are diagonalisable to have positive diagonal entries.

Contrary to $\U$, the $\pos$ map does not preserve the purity $\Tr{\rho^2}$ because the matrix on a real vector representation does not preserve the Euclidean norm, but it still preserves the trace of any density operator. As such, it can be interpreted as an irreversible process in quantum dynamics. The characteristics of $\U$ and $\pos$ together will be employed in the analysis of quantum dynamical maps in subsequent sections.

\section{Decomposition of Dynamical Matrix}\label{sec:decom}
To demonstrate the idea of the unitary-scaling decomposition, we consider the Lindblad formulation for the dynamical maps. In this case, a dynamical map is given by a continuous mapping $t\mapsto\Phi_t$, where $t\in\left[0,\infty\right)$ parametrises time of the dynamics, and the map $\Phi_t$ acts on either the state space (Schr\"odinger picture) or on the operator space (Heisenberg picture).
The original definition of a Lindblad map in the Heisenberg picture as a semigroup on a $C^*-$algebra is provided in \ref{appen:Lind}. In this work, we consider the Schr\"odinger picture and employ the dual-formalism of the map on the state space as we are primarily concerned with the evolution of a quantum state. In this sense, $\Phi_t$ acting on the state space $\st_d$ can be written formally in the exponential form as $\Phi_t = e^{t\Lind},$ where
	\begin{eqnarray}
		\Lind(\rho) &:=& -i\left[\op{H},\rho\right] + \sum_\alpha\diss_{h_\alpha}(\rho),
		\label{eq:Lindblad}\\
		\diss_{h_\alpha}(\rho) &:=& h_\alpha\rho h^\dagger_\alpha-\frac{1}{2}\left( h_\alpha h^\dagger_\alpha\rho +  \rho h_\alpha h^\dagger_\alpha \right).
		\label{eq:disspart}
	\end{eqnarray}
Here $\op{H}$ is a self-adjoint operator in $\Banach(\Hilbert),$ the set of bounded operators on  $\Hilbert;$ and $h_\alpha$ is an operator in $\Banach(\Hilbert)$ satisfying $\sum_\alpha h^\dagger_\alpha h_\alpha \in \Banach(\Hilbert),$ where the sum is taken over some countable set. 

There exists a dynamical matrix $\M_t$ in $\Mat_{d^2-1}(\Real)$ corresponding to $\Phi_t$ as $\Phi_t=\varphi[\M_t],$ as in Eq. \eqref{eq:counterpartmap}. We obtain a vector equation $\X_t = \M_t\cdot\X$, where $\X_t:=\X(\rho_t)$, $\X:=\X(\rho),$ rather than the dynamical equation $\rho_t=\Phi_t(\rho).$ Because the mapping $t \mapsto \Phi_t$ is bounded, hence continuous, it can be viewed as a trajectory on the vector space $\Real^{d^2-1}$ mapped out by $\M_t.$ As stated, our main objective is to analyse the role of the unitary-scaling decomposition of a unital dynamical map. To that end, we will write the Lindblad map and its real-vector representation in such the decomposition. First, we state the polar decomposition lemma whose proof can be found in \cite{Horn}.
\begin{thm}{(Polar Decomposition)}
	Any $\op{A}\in\Mat_d$ can be written as $\op{A} = \op{P}\op{U},$ where $\op{P}$ is positive semidefinite, and $\op{U}$ is unitary. The matrix $\op{P}$ is always uniquely determined as $\op{P}=(\op{A}\op{A}^*)^{1/2}.$ Additionally, if $\op{A}$ is non-singular, then $\op{U}$ is uniquely determined as $\op{U}\equiv\op{P}^{-1}\op{A}.$ If $\op{A}$ is real, then $\op{P}$ and $\op{U}$ may be taken to be real.
\end{thm}

In our setting, the matrix $\M\in\Mat_d(\Real)$ has a polar decomposition $\M = \pp\V$ for some positive symmetric matrix $\pp,$ and isometric orthogonal matrix $\V$ (analogous to $\op{P}$ and $\op{U}$ in the theorem above). When $\M$ is normal, it also has the form $\M=\V\pp$ \cite{Horn}. By construction, the dynamical matrix $\M_t$ of the Lindblad map with normal and unital properties has the polar decomposition $\M_t = \V_t\pp_t = \pp_t\V_t,$ leading to the same decomposition of their induced maps; namely, $\Phi_t=\varphi[\M_t] = \varphi[\V_t]\circ\varphi[\pp_t] = \varphi[\pp_t]\circ\varphi[\V_t].$ 

As a transformation, the isometric orthogonal part $\V_t$ can be identified as a rotation in $\Real^{d^2-1},$ while the positive part $\pp_t$ as a scaling map when $\M_t$ is a contraction map, as in our case. For this reason, the induced maps $\varphi\left[\rot_t\right]$ and $\varphi\left[\scale_t\right]$ will be treated as reversible and irreversible processes, respectively. We remark that the interpretation of the unitary mapping on a finite dimensional space as a rotational transformation on the real vector space is common in the field of information geometry \cite{geobook}. Similarly, $\varphi\left[\scale_t\right]$ can reflect the dissipative behaviour, e.g. distilling heat from work. This will be clear when we consider a close relationship between the rotation and scaling parts in Section \ref{subsec:semigroup}.

\section{Subspace Separation of the Dynamics}\label{sec:observe}
In this section, we present our results from the consequences of the unitary-scaling decomposition of the dynamical matrix arisen from the normal and Markovian properties of the unital Lindblad map. In essence, the dynamical matrix, and its rotation and scaling parts are simultaneously block-diagonalisable and share the same eigensubspaces. We begin by investigating the structure of the rotation matrix as it leads to the condition of the form on the scaling part. A special case when the scaling part of the dynamical map is isotropic is treated in Section \ref{subsec:markov}. An example in a qubit case will be demonstrated in Section \ref{sec:qubit} to discuss the insights from our results, including an anisotropic case. 

\subsection{Scaling Part of Normal Dynamical Matrix}
It has been already shown that the normality property is a sufficient condition for relaxation with respect to the Hilbert-Schmidt norm \cite{WangSchirmer2009}, and, the dynamics would be relaxed within this norm if and only if it is unital \cite{PerezEA2006}. 
In this sense, even though there exist non-normal (usually not relaxed) dynamical Lindblad generators, many important systems in physics lie in the normal class. Hence, essentially without loss generality, it suffices to consider only a case of normal dynamical matrix for the unital evolution.

For a normal matrix $\M_t,$ it follows that $\M_t = \rot_t\scale_t = \scale_t\rot_t,$ where $\rot_t$ and $\scale_t$ denote the associated rotation and scaling parts, and $t$ is a time parameter. Consider a rotation matrix $\rot$ as an element of  $\SO(n)$, the set of special orthogonal matrices \cite{rossmann}.
\begin{lem}
	Every element $\rot \in \SO(n)$ is conjugate to a block-diagonal matrix; 
	\begin{eqnarray*}
		\rot \cong diag\left( r_1, r_2, \cdots, r_m \right) &{\hspace*{2em} if \hspace*{2em}} n=2m;\\
		\rot \cong diag\left( r_1, r_2, \cdots, r_m,\iden_1\right) &{\hspace*{2em}  if  \hspace*{2em}} n=2m+1.
	\end{eqnarray*}
	For each $k = 1, 2, \ldots, m$, we can identify {$r_k = \left(\begin{array}{cc}
	\cos\theta_k & -\sin\theta_k \\ 
	\sin\theta_k & \cos\theta_k
	\end{array}  \right)$} for some angle $\theta_k$, and $\iden_1$ is the $1\times 1$ identity block.
	\label{lem:canoform}
\end{lem}

\begin{coll}
		In any two dimensional eigensubspace, a positive symmetric matrix $s \in \Mat_2(\Real)$ commuting with a rotation matrix $r\in\SO(2)$ must be $\alpha\iden_2$, where $\alpha$ is a positive real number, unless $r$ is itself the identity matrix $\iden_2$ in $\Mat_2(\Real).$ \label{rem:item:commuteYiden}
\end{coll}
The claim easily follows from considering a matrix $s = \left(\begin{array}{cc}
	a & b \\ 
	b & c
	\end{array}  \right)$, and solving for the entries from the condition $s r=rs.$

\begin{prop}
	Any scaling matrix $\scale \in \Mat_n(\Real)$ that commutes with a rotation matrix $\rot \in \Mat_n(\Real)$ is conjugate to a block-diagonal matrix; 
	\begin{eqnarray*}
		\scale &\cong& diag\left( e^{-\lambda_1}\iden_2, e^{-\lambda_2}\iden_2, \cdots, e^{-\lambda_m}\iden_2 \right) 
		 {\hspace*{3.6em}  if  \hspace*{1em}} n=2m,\\
		\scale &\cong& diag\left( e^{-\lambda_1}\iden_2, e^{-\lambda_2}\iden_2, \cdots, e^{-\lambda_m}\iden_2, e^{-\lambda_{m+1}}\iden_1\right) {\hspace*{1em}  if  \hspace*{1em}} n=2m+1.
	\end{eqnarray*}
	Here $\lambda_k$ is a non-negative real number for each $k=1,2,\ldots,m+1$.
	\label{prop:scaleplane}
\end{prop}

A proof of the above proposition is provided in \ref{appen:proof}. By the block-diagonal form of the rotation and scaling matrices, the normal dynamical matrix $\M_t=\rot_t\scale_t$ can be arranged into the same block diagonal form by the same similarity transformation. It is convenient to employ this form for the dynamical matrix $\M_t.$ We remark here that this block-diagonal form arises strictly from the setting of the dynamical matrix, while only the condition of the normality property is employed. Thus, the same analysis can be adopted to the generalised class of dynamical maps which possesses the normality property. 

\subsection{Markovianity and Isotropic Scaling}\label{subsec:markov}

For the dynamical matrix $\M_t,$ we call it Markovian whenever $\M_{t+s} = \M_t\M_s$ for all $t,s > 0.$ 
With the Markovian property, the dynamical matrix $\M_t$ can be written as $\displaystyle\M_t=\prod_{i=1}^N\M_{\tau_i},$ with the time duration $\tau_i>0$ and $\displaystyle\sum_{i=1}^N\tau_i=t.$
If all $\scale_{\tau_i}$ are isotropic, we obtain that
	\begin{equation}
		\rot_t\scale_t	=	\prod_{i=1}^N\rot_{\tau_i}\scale_{\tau_i} =	\left(\prod_{i=1}^N\rot_{\tau_i}\right) \left(\prod_{i=1}^N\scale_{\tau_i}\right).
	\end{equation}
In addition, since $\rot_t$ is a group with parameter $t$, we can deduce that $\scale_t = \prod_{i=1}^N\scale_{\tau_i}$ obeys the addition property of a semigroup. Therefore, with the isotropic scaling,
	\begin{equation}
		e^{-\lambda(t)}=\exp{\left(-\sum_{i=1}^N\lambda({\tau_i})\right)},			\label{eq:lambdaTruncate}	
	\end{equation}
where $\lambda(t)$ denotes the scaling parameter of the map at time $t$ (which is indeed the time duration in the Markovian case). Since the partition $\{\tau_i\}_{i=1}^N$ of the time duration $t$ is arbitrary, we can conclude that $\lambda(t)$ must be a linear function of $t,$ say $\lambda(t)=\gamma t$, where $\gamma$ is a positive real number. This follows from the continuity property of the Lindbladd map. We will later discuss the characteristics of the scaling parameter function $\lambda_k(t)$ for each block in the non-isotropic case in Section \ref{subsec:semigroup}. For now, we consider the consequences of the Markovian property and isotropic scaling on qubit systems.

\subsection{Qubit Systems}\label{sec:qubit}
\begin{figure}
\begin{center}
	\includegraphics[scale=0.6]{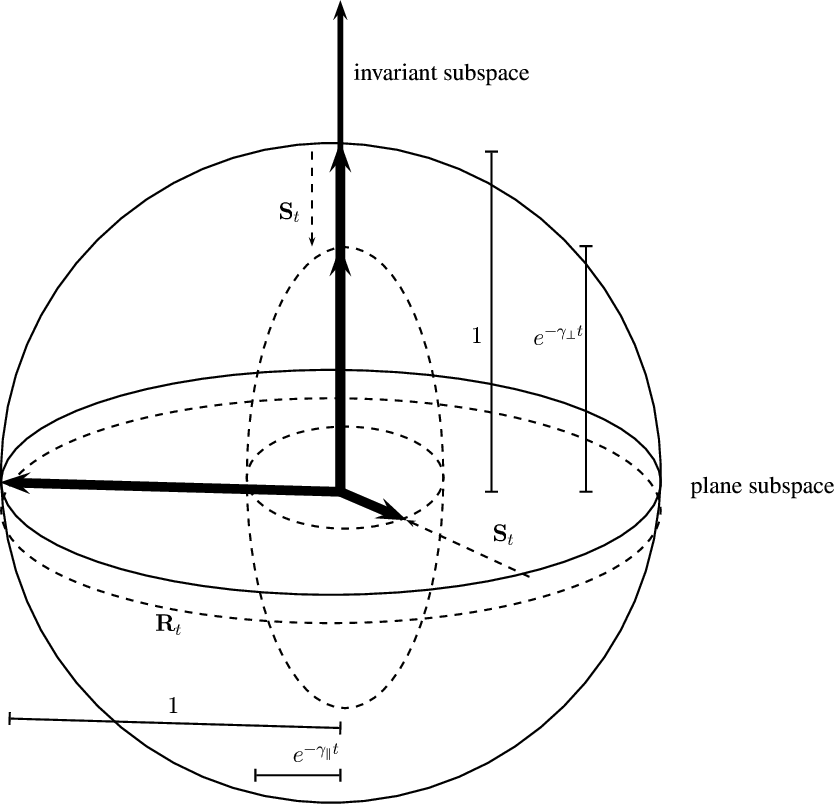}
\end{center}
\caption{Geometric representation of the Bloch ball for the a qubit system. (in plane) A plane subspace where the component of a given vector is rotated by $\rot_t$ and scaled by $\scale_t$. (vertical component) The invariant subspace where the component remains the same via rotation, but still has a change in radius by the effects from the scaling map.}
\label{fig:domainreduce}
\end{figure}

In a qubit system, we can write the quantum state $\rho$ as 
	\[ \rho=\frac{1}{2}\iden_2 + \sum_{\alpha=1}^3 f_\alpha x_\alpha(\rho), \]
where $\{ \frac{1}{2}\iden_2, f_1, f_2, f_3 \}$ denotes the basis set which comprises Pauli's matrices. In this case, Lemma \ref{lem:canoform} is exactly the Euler's theorem, where the block-diagonal form can be written as 
\[\rot^D_t = diag(r_1(t),\iden_1).\] 
That is, any rotation matrix in $\Real^3$ rotates components of a real vector representation of an initial state within a certain plane and leaves the other component invariant. We will refer to the projection of any vector onto this plane as a plane component, and the remaining orthogonal block $\iden_1$ an invariant (or orthogonal) component. Indeed, the invariant component is interpreted as the axis of rotation. Moreover, the two invariant vectors of the rotation in the ball (the pole vectors) are the real vector representations of projections (or pure density matrices in $\st_d$) corresponding to the eigenvectors of the Hamiltonian $\Hamil,$ which is a generator of $\U_t.$ The scaling matrix $\scale_t$ is reduced to the form
\[\scale_t = diag (e^{-\gamma_{\parallel} t}\iden_2, e^{-\gamma_{\perp} t}), \]
where the exponents $\lambda_1(t):=-\gamma_{\parallel}t$ and $\lambda_2(t):=-\gamma_{\perp}t$ are the scaling parameters for the plane and invariant subspaces with the decay rates $\gamma_{\parallel}$ and $\gamma_{\perp}$, respectively. Note that the linearity arises from the Markovian property of the dynamics. Here, the scaling of a vector is different between the in-plane and the orthogonal components. Consequently, the image of the dynamical matrix is transformed from the spherical ball to a prolate spheroid when $\gamma_{\parallel} > \gamma_{\perp}$, and an oblate one when $\gamma_{\parallel} < \gamma_{\perp},$ or a smaller spherical ball when the scaling is isotropic; as illustrated in Figure \ref{fig:domainreduce}.

It should be remarked that similar results for the qubit case were extensively analysed in literature with different approaches; for examples, the canonical form for the master equations and the measure of decoherence \cite{PhysRevA.89.042120}, the studies of Davies' maps on qubits \cite{ROGA2010}, the analysis of minimal entropy of evolved states \cite{KingRuskai2001}, and the analysis of completely-positive trace-preserving maps on $\Mat_2(\complex)$ \cite{BETHRUSKAI2002}. When a translation is also taken into account,  the decomposition yields a product of scaling and rotation, followed by translation in the homogeneous coordinates. It has been shown that, for a unital map, the entropy change, related to only the scaling part, is identified as an entropy production \cite{KingRuskai2001}. When the translation is present, the exchange entropy also arises \cite{Alicki}. 

\section{Entropy Change Profile in the Dynamics}\label{sec:entropy}
In this section, we will consider the entropy change along the dynamics and investigate the role of the scaling parameters in such change. For technical simplicity, we employ the linear entropy to illustrate the effect, which is given for the isotropic case first and will be extended to the anisotropic case thereafter. Again, the geometrical picture will be exemplified in the qubit system framework.

\subsection{Entropy Change for Isotropic Scaling}
\begin{figure}[ht]
\begin{center}
\includegraphics[scale=0.65]{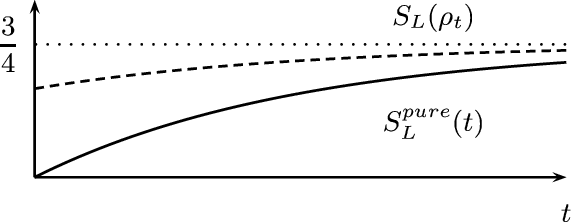}
\end{center}
\caption{The solid line shows the linear entropy $s^{pure}(t)$ along the Lindblad dynamics when the initial state is a pure state for the case that the scaling part $\scale_t$ is isotropic. The dashed line is the linear entropy for the same initial state with $0.5$ purity. Here, we set the scaling parameter $\gamma=1,$ and dimension $d=4$ (e.g. two-qubit systems).}
\label{fig:exEntrochange}
\end{figure}

For the dynamical matrix of isotropic scaling, one can obtain the linear entropy in the form
	\begin{equation}
		\ls(\rho_t)	= s^{pure}(t) + e^{-2\gamma t}\ls(\rho),\label{eq:isoEnt}
	\end{equation}	
where
	\begin{equation}
		s^{pure}(t) = \frac{d-1}{d}\left(1 - e^{-2\gamma t} \right)
		\label{eq:exEntrochange}
	\end{equation}
denotes the linear entropy in the case that the initial state is a pure state, and $\ls(\rho):=\left( 1 - \Tr{\rho^2} \right)$ denotes the linear entropy of the actual initial state of the dynamics. This expression shows that the Lindblad or Markovian dynamics yield the linear entropy increasing with time and possessing an asymptote or bound by a number $\frac{d-1}{d}$ as exemplified in Figure \ref{fig:exEntrochange}. We point out again that the scaling parameter is in one-to-one correspondence with the entropy. This allows us to consider the scaling parameter as a quantification of non-adiabaticity, and likewise verify that entropy is a dynamical parameter, as commonly believed. 
More importantly, it also shows that the entropy change contains only the contributions from the scaling part, so we can essentially say that the scaling part is an irreversible action for a unital map. 

\subsection{Semigroup Properties and Anisotropic Scaling}\label{subsec:semigroup}

Now we discuss the semigroup properties in the Markovian dynamics with anisotropic scaling. Suppose we can write $\M_t=e^{t\matL}$ for some matrix $\matL$, which denotes the generator of the semigroup $\M_t$ and corresponds to the generator of the semigroup $\Phi_t$. The correspondence between $\M_t = e^{t\matL}$ and $\Phi_t$ follows from multiplicativity of the mapping $\varphi$ and by means of the density of the set of polynomials in $\Her_d$ together with the spectrum theorem \cite{mathlang}. 

From the construction, we know that all the eigenvectors are independent of time $t,$ while the eigenvalues are not. These properties are inherited to $\K\rot_t\K^{-1},$ where the matrix $\K$ is a similarity transformation matrix on $\Real^{d^2-1}$ from the coordinates in the basis $\X(\totmixed +f_\alpha)$ to the components in the eigensubspace of $\rot_t$; hence, $\K$ is orthogonal and also independent of time $t.$ Since $\scale_t$ commutes with $\rot_t$ by the normal property of $\M_t$, we obtain
	\begin{equation}
		\M_t = \K\rot^D_t\scale^D_t\K^
{-1},
	\end{equation}
where $\rot^D_t$ and $\scale^D_t$ denote the block-diagonal forms of $\rot_t$ and $\scale_t,$ respectively. It is sufficient to consider the action only in each subspace described by the block in $\rot^D_t$ or $\scale^D_t.$ Let $m_k(t),$ $r_k(t)$ and $s_k(t)$ be the $k^{th}$ block of the matrices $\rot^D_t\scale^D_t,$ $\rot^D_t$ and $\scale^D_t,$ respectively. Without loss of generality, we assume that $d^2-1$ is even; otherwise, there is additionally the $1\times 1$ identity block which can be treated separately and trivially. Then,
\[
	m_k(t) := r_k(t)s_k(t) = e^{-\lambda_k(t)}{\left(\begin{array}{cc}
	\cos\theta_k(t) & -\sin\theta_k(t) \\ 
	\sin\theta_k(t) & \cos\theta_k(t)
	\end{array}  \right)},
	\]
where $\lambda_k(t)$ and $\theta_k(t)$ are functions of $t$ related to the scaling parameter of $\scale_t$ and the angle of rotation of $\rot_t$ in the $k^{th}$ eigenblock, respectively. Furthermore, the scaling matrix $s_k(t)$ is clearly isotropic within the eigenblock, leading to the following statement whose proof can be found in \ref{appen:proof}.

\begin{prop}
The functions $\lambda_k(t)$ and $\theta_k(t)$ are either simultaneously additive in $t$ or simultaneously non-additive in $t$ for all $k=1,\ldots,m$ with $2m=d^2-1$. 
	\label{prop:bothlinearOrnot}
\end{prop}

By the similar reasoning as in Section \ref{subsec:markov}, since the partition of  time interval $\left[0,t\right]$ into $\{\tau_i\}_{i=1}^{N}$ is arbitrary, together with continuity for the Lindblad map, the additivity is equivalent to the linearity. Thus, $\lambda_k(t)$ and $\theta_k(t)$ are either both simultaneously linear in $t,$ or they are simultaneously not linear in $t$. Unlike the isotropic condition, Proposition \ref{prop:bothlinearOrnot} together with the fact that $r_k(t)$ preserves the 2-dimensional Euclidean norm, implies that
	\begin{equation}
		\ls(\rho_t) = \frac{d-1}{d} - \sum_{k=1}^m e^{-2\gamma_k t}\ENORM{\X_k(\rho)}^2.
		\label{eq:exEnt_nonISO}
	\end{equation}
If $\gamma_k=\gamma$ for all $k = 1,\ldots,m,$ Eq. \eqref{eq:isoEnt} for the isotropic case is recovered as expected. From Eq. \eqref{eq:exEnt_nonISO}, the linear entropy is expressed as a weighted sum of the exponential-decay functions in each component, where the weight in the $k^{th}$ subspace is equal to $\ENORM{\X_k(\rho)}^2$ of the initial state $\rho.$ It again reflects the properties of increasing in time and possessing an asymptote for the Markovian dynamics. Also, it should be emphasised that from Eq. \eqref{eq:exEnt_nonISO}, not only the scaling parameter $\gamma_k$ corresponding to each subspace, but also the mass $\ENORM{\X_k(\rho)}^2$ therein the subspace from the initial state affect the dissipative behaviour of the system.  As an example, when the initial state $\rho$ has the real vector representation lying in only one of the subspaces, say the $k^{th}$ subspace, but all other components are null. That is, $\ENORM{\X_k(\rho)}^2 \neq 0$ and $\ENORM{\X_l(\rho)}^2 = 0$ for all $l\neq k.$ Then, 
	\begin{equation}
		 \ls(\rho_t) = \frac{d-1}{d} - e^{-2\gamma_k t}\ENORM{\X_k(\rho)}^2,
		 \label{eq:linEntEach}
	\end{equation}
retrieving the isotropic case. However, if the real vector representation of the initial state $\X(\rho)$ has multiple components, then the change of the linear entropy is constituted by more than one rate. In summary, the characteristics of map, the initial state $\rho$ and the relationship among them affect the change of entropy.

\subsection{Qubit Systems}\label{subsec:EntChange}
\begin{figure}
	\begin{center}
	\includegraphics[scale=0.7]{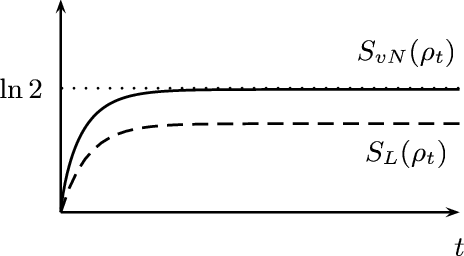}
\end{center}
	\caption{The solid line shows the change of von Neumann entropy in time for dynamical map with isotropic scaling with $\gamma=0.5$ and from a pure initial state ($\Tr{\rho^2}=1$) while the dashed line represents the linear entropy for the same situation. }
	\label{fig:QbitvSIso}	
\end{figure}

The expression of the linear entropy in Eq. \eqref{eq:exEnt_nonISO} for $d=2$,
	\begin{equation}
		\ls(\rho_t) = \frac{1}{2} - \left( e^{-2\gamma_{\parallel}t}\ENORM{\X_{\parallel}(\rho)}^2 + e^{-2\gamma_{\perp}t}x^2_{\perp}(\rho)\right),
	\end{equation}
is reduced to $\ls(\rho_t) = \frac{1}{2} - e^{-2\gamma t }\ENORM{\X(\rho)}^2$ when the scaling part is isotropic with $\gamma_{\parallel} = \gamma_{\perp} =\gamma.$ Thus, the interpretation given in the previous section for this change is also applied to this situation. For this particular case, the linear entropy can explicitly demonstrate the behaviour of the dynamics from the geometric interpretation of the state along the dynamics. Although we employ the linear entropy because it is practically convenient to calculate, there is another insight from this expression. Since the purity or the radius of the Bloch vector is used in result exploration in many experiments, the connection between measurement results and the characterisation of the dynamical map can be made. For instance, the rate of change in the entropy along the dynamics can be expected directly from the invariant component, which should reflect the stationary situation in the dynamics. Moreover, in a qubit system, the von Neumann entropy \cite{Barnett} can be explicitly expressed as 
	\begin{eqnarray}
		\vs(\rho) &=& -p_1\ln{p_1} - p_2\ln{p_2},
	\end{eqnarray}
where $p_1$ and $p_2$ are eigenvalues of $\rho$ given by $p_1 = \frac{1}{2}\left(1+r\right)$ and $p_2 = \frac{1}{2}\left(1- r\right).$ The radius $r$ of the state $\rho_t$ is indeed the same as the vector norm in the linear entropy, i.e. 
\[ r:=\ENORM{\X_t(\rho)} = \sqrt{e^{-2\gamma_{\parallel}t}\ENORM{\X_{\parallel}(\rho)}^2 + e^{-2\gamma
_{\perp}t}x^2_{\perp}(\rho)}. \]
By this form, the effects of the initial preparation in the selection of the rate of dissipation is not as obvious as by the linear entropy, but it is still clear that the change of the von Neumann entropy depends on the weights $\ENORM{\X_{\parallel}(\rho)}^2$ and $x^2_{\perp}(\rho)$ via the radius of the vector representation. In particular, when the scaling matrix is isotropic, 
	\begin{eqnarray}
	\hspace{-2cm} 	\vs(\rho) &=&  -\frac{1}{2}\left(1+e^{-\gamma t}r_0\right)\ln{\frac{1}{2}\left(1+e^{-\gamma t}r_0\right)} -\frac{1}{2}\left(1-e^{-\gamma t}r_0\right)\ln{\frac{1}{2}\left(1-e^{-\gamma t}r_0\right)},
	\end{eqnarray}
where $r_0 = \ENORM{\X},$ and $\gamma$ is a scaling parameter. As expected, it also reflects the increasing and possessing asymptote properties of the entropy; as illustrated in Figure \ref{fig:QbitvSIso}.

\section{Conclusion}\label{sec:discussion}
We investigate the  decomposition of a dynamical map of the unital Lindblad type, which is used in quantum dynamics of a finite dimensional open quantum system, into two distinct types of mapping on the space of quantum states represented by the density matrices. Based on the real-polar decomposition of a matrix, the formulation and decomposition of a dynamical map employ the interplay between the density matrix and the Bloch representations of the state. One component of the decomposed dynamical map corresponds to unitary evolution or reversible or coherent behaviours represented by a rotation matrix, while the other to irreversible characteristics or dissipative or decoherent behaviours represented by a scaling matrix. 

Under the normality condition, we show that the scaling parameter is simply a linear function of time, which is also in one-to-one correspondence with the entropy. As expected, the behaviour of the Markovian dynamics and the change of linear entropy or purity, as an indicator of dissipative behaviours, increases in time and possesses an asymptote. 

More importantly, the issues of initial-state dependence of dissipative behaviours, and the role of eigensubspace partitioning of the dynamical matrix are discussed. In particular, the rate of change of the linear entropy depends on the structure of the scaling part of the dynamical matrix, where the initial state plays an important role in this rate of change.  Specifically, the linear entropy is expressed as a weighted sum of the exponential-decay functions in each scaling component, where the weight is equal to $\ENORM{\X_k(\rho)}^2$ of the initial state $\rho$ in the subspace. 

From the unitary-scaling decomposition, it is evident that the dissipative behaviour of an open quantum system may be considered in parallel with the coherent behaviour. One can set the unitary dynamical map as a centred map and the scaling map as the deviation from the unitary map. In such ways, the use of linear entropy, which is adopted as an indicator of the dissipative behaviour and as another dynamical parameter becomes a promising tool in understanding evolution of an open quantum system.

\section*{Acknowledgement}
We are grateful to Dr. Kavan Modi and Dr. Felix Pollock of Monash University, Australia, for valuable discussion and guidance. FS also thanks Sri-Trang Thong Scholarship, Faculty of Science, Mahidol University, for financial support to study at the University's Department of Physics.

\bibliographystyle{model1-num-names}
\bibliography{manuscript.bbl}

\appendix

\section{Lindblad Dynamical Map}\label{appen:Lind}
\setcounter{section}{1}
In the original work of Lindblad \cite{Lindblad}, the map is expressed in the Heisenberg picture using the $C^*$-algebra framework that a dynamical map acts on the set of quantum operators. Let $\Phi_t$ denote a dynamical map on $\st_d,$ and $\Phi_t^*$ its corresponding dual map on the operator space $\Mat_d(\complex)$, which are related by the relation
$
	\Tr{\rho\Phi_t^*(\op{a})}=\Tr{\Phi_t(\rho)\op{a}},
$		
for all $\rho\in\st_d$ and all operators $\op{a}\in\Mat_d(\complex)$. We call $\Phi_t^*$ a Lindblad dynamical map if its satisfies the following conditions.
	\begin{enumerate}[(L1)] 
		\item $\Phi_t^*\in CP_\sigma(\Hilbert),$ \label{L1}
		\item $\Phi_t^*(\iden_{\Banach(\Hilbert)})=\iden_{\Banach(\Hilbert)},$\label{L2}
		\item $\Phi_t^*\Phi_s^* = \Phi_{t+s}^*,$ \label{L3}
		\item $\lim_{t\searrow 0} \norm{\Phi_t^* - \iden_{\Banach(\Banach(\Hilbert))}} = 0,$ \label{L4}
	\end{enumerate}
where $CP_\sigma(\Hilbert)$ denotes the set of all completely positive maps in the space of bounded operators $\Banach(\Hilbert),$ or equivalently $\Mat_d(\complex).$ The map $\Phi_t^*$ with its tensor extension $\Phi_t^*\otimes\iden_n : \Banach(\Hilbert)\otimes\Mat_n(\complex) \rightarrow \Banach(\Hilbert)\otimes\Mat_n(\complex)$ is positive for all $n=1,2,\ldots$ (see \cite{Lindblad,Alicki,grech}).  The conditions (L\ref{L1}) and (L\ref{L2}) correspond to (C\ref{as:PTP}) and also yield the condition (C\ref{as:contract}). (L\ref{L3}) is the Markovian property. 

\section{Unitary-Scaling and Translation-Scaling-Unitary Decompositions}\label{appen:SR-TSR}

Since we are interested in separating a purely decoherence component and a coherence component of a given dynamical map, it is useful to comment on general ideas of the dynamical map decomposition. In fact, the employment of the polar decomposition on the matrix representation on the Hilbert-Schmidt space of the qubit case to investigate the region of maximum entropy has been done before \cite{KingRuskai2001}. In that work, it was pointed out that the main advantage of performing such composition on the Hilbert-Schmidt space, instead of directly on arbitrary bases of $d^2$ dimensional vector space, is that one could obtain a clearer geometric picture of the dynamics, as well as the structure of the state along the dynamics, including asymptotic characterisation. 

As an example, let's consider the dynamics generator is time independent. Recall the expression
	\begin{equation}
		\rho = \sum_{\alpha=0}^{d^2-1}x_{\alpha}f_{\alpha}.
	\end{equation}
Then, the dynamical map $\Phi$ can be represented as a linear map $\op{M}$ on $\Real^{d^2}$ in the homogeneous coordinates as
	\begin{equation}
		\op{M}:={\left(\begin{array}{cc}
		1 & \vec{0}^{T} \\ 
		\vec{c} & \op{M}_-
		\end{array} \right)}. \label{eq:cM}
	\end{equation}
where $\vec{0}^{T}$ is a transposed zero vector in $d^2-1$ dimensions; and $\op{M}_-$ is a unital map in $d^2-1$ real vector space. Clearly, the matrix elements in this basis can be obtained from $\op{M}_{\alpha\beta}=\HS{f_\alpha}{\op{M}\cdot\X({f_\beta})}.$ 
If the dynamics are described by $\dot{\rho_t}=\Lind(\rho),$ by the orthogonality of $F,$ it follows that
	\begin{equation}
		\frac{d}{dt}\X_t = \genL\cdot\X_t + \vec{\ell},
	\end{equation}
where $\genL_{\alpha\beta}=\HS{f_\alpha}{\Lind({f_\beta})}$ for $\alpha \neq 0$ and $\beta \neq 0$. Here $\ell_\alpha=\HS{\Lind(\iden_d)}{f_\alpha}$ for $\alpha\neq 0$ contains information of the translation. Let $\X_0$ be the initial vector representation of the state; then, one obtains
	\begin{equation}
		\X_t = e^{t\genL}\cdot\X_0 + \vec{c}_t,
	\end{equation}
where the translation vector is 
\begin{equation}
\vec{c}_t = \left(e^{t\genL}-\iden_{\Real^{d^2-1}}\right)\cdot\genL^{-1}\cdot\vec{\ell},
\end{equation} 
and $\iden_{\Real^{d^2-1}}$ is the identity matrix acting on $\Real^{d^2-1}.$ One can see that for unital map, e.g. $\Lind(\iden_d) = 0,$ the translation vanishes, yielding the expression in Eq. \eqref{eq:counterpartmap}. Next we would like to discuss the decomposition in two different ways.

\subsection{Translation-Scaling-Rotation (TSR) Decomposition}
From Eq. \eqref{eq:cM}, in a finite dimensional system, it can be said that a dynamical map can be decomposed into a translation map and a unital map in some hyperspace. In this sense, one can see that the structure of the map can be considered as combination of a map acted on two parts of the state, i.e. the tracial part spanned by $\iden_d/d$ and the traceless part (or the Bloch region). While the unital component of the map concerns only the rearrangement in the $\Real^{d^2-1}$, the translation part, whose transformation on $\Real^{d^2}$ is denoted by $\op{T}$, will represent the transfer of the tracial part to the traceless one. For a valid dynamical map in the Shr\"odinger picture, the reverse transformation from the traceless part to the tracial one is zero by the normalisation condition.

According to the unitary-scaling decomposition, the unital part $\op{M}_-$ can also be decomposed into scaling matrix $\op{S}_-$ and the rotation $\op{R}_-$. Let $\op{S}$ and $\op{R}$ denote their extensions in $\Real^{d^2}$. Then, one can write 
    \begin{equation}
        \op{M} = \op{T}\circ\op{S}\circ\op{R},
    \end{equation}
which is termed the translation-scaling-rotation (TSR) decomposition. We remark that only $\op{T}$ and $\op{S}$ can contribute the entropy change. However, it is shown that the thermodynamic entropy can be affected only by the translation part $\op{T}$, while the entanglement entropy can be contributed by both $\op{T}$ and $\op{S}$ \cite{Alicki}. This result verifies intuition since the totally mixed state (aka an infinite temperature state) is an asymptotically fixed point for a unital map, but not so for a non-unital map. Indeed, this state could be mapped to another Gibb's state of finite temperature instead. Hence, in addition to the geometric picture interpretation of quantum dynamics, another advantage of the only unitary-scaling decomposition, when possible, is that one can obtain insights about different thermodynamic characteristics, which would be useful in describing the thermodynamic behaviour of the dynamical map. 

\subsection{Scaling-Unitary (SR) Decomposition}
For a finite dimensional Hilbert space, a dynamical map $\Phi$ can be represented as a matrix $\op{M}$ in an arbitrary basis. Although the representation on $\Real^{d^2-1}$ is not linear for non-unital case due to the existence of the translation part, it is linear in $\Real^{d^2}$, so one can directly apply the polar decomposition to obtain $\op{M} = \tilde{\op{S}}\circ\tilde{\op{R}}$, where $\tilde{\op{S}}$ is a positive symmetric matrix and $\tilde{\op{R}}$ is a unitary matrix. Unlike in the Hilbert-Schmidt representation, the matrix is not taken over the field of real numbers, instead the complex ones. In this approach, one can see that the entropy is contributed by the symmetric part $\tilde{\op{S}}$, whereas $\tilde{\op{R}}$ only corresponds the change in phase. The advantage of this decompostition is that the irreversible characteristics, such as the entropy change, can be determined only by $\tilde{\op{S}}$. However, one will lose the geometric picture of the quantum dynamics and also the thermodynamic characterisation.

\section{Proofs of Propositions \ref{prop:scaleplane} and \ref{prop:bothlinearOrnot}}\label{appen:proof}
\begin{proof}{(Proof of Proposition \ref{prop:scaleplane})}
	Since $\scale$ and $\rot$ commute, they share the same eigenbases, and both are simultaneously block-diagonalisable. By employing Lemma \ref{lem:canoform}, so the scaling matrix $\scale$ is conjugate to a block-diagonal matrix;
	\begin{eqnarray}
		\scale \cong diag\left( s_1, s_2, \cdots, s_m \right) &{\hspace*{2.2em}  if  \hspace*{1em}} n=2m;\\
		\scale \cong diag\left( s_1, s_2, \cdots, s_m, \alpha_{m+1}\iden_1\right) &{\hspace*{2em}  if  \hspace*{1em}} n=2m+1. 
	\end{eqnarray}
where $s_k$ is a $2\times 2$ positive symmetric block for $k=1,\ldots,m$. Since the commutativity will apply to the $k^{th}$ eigenspace the blocks $r_k$ and $s_k$ also commute. From Corollary \ref{rem:item:commuteYiden}, it follows that $s_k = \alpha_k \iden_2$ for some $\alpha_k >0$ and $k = 1, 2, \ldots, m$. As the scaling map in our case is a contraction, there exists $\lambda_k \ge 0$ such that $\alpha_k = e^{-\lambda_k}$, for $k = 1, 2, \ldots, m+1$.
\end{proof}

\begin{proof}{(Proof of Proposition \ref{prop:bothlinearOrnot})\\}
From the Markovian property, we have $m_k(t_1+t_2)=m_k(t_1)m_k(t_2)$: 
	\begin{eqnarray}	
		\hspace{-2cm} m_k(t_1+t_2)	 &=&	e^{-\lambda_k(t_1+t_2)}{\left(\begin{array}{cc}
	\cos\theta_k(t_1+t_2) & -\sin\theta_k(t_1+t_2) \\ 
	\sin\theta_k(t_1+t_2) & \cos\theta_k(t_1+t_2)
	\end{array}  \right)}.\\
		\hspace{-2cm}		 m_k(t_1)m_k(t_2) 	&=&	e^{-\left[\lambda_k(t_1)+\lambda_k(t_2)\right]}{\left(\begin{array}{cc}
	\cos\theta_k(t_1) & -\sin\theta_k(t_1) \\ 
	\sin\theta_k(t_1) & \cos\theta_k(t_1)
	\end{array}  \right)}{\left(\begin{array}{cc}
	\cos\theta_k(t_2) & -\sin\theta_k(t_2) \\ 
	\sin\theta_k(t_2) & \cos\theta_k(t_2)
	\end{array}  \right)}\nonumber\\
		\hspace{-2cm}		&=&e^{-\left[\lambda_k(t_1)+\lambda_k(t_2)\right]} {\left(\begin{array}{cc}
	\cos\left[\theta_k(t_1)+\theta_k(t_2)\right] & -\sin\left[\theta_k(t_1)+\theta_k(t_2)\right] \\ 
	\sin\left[\theta_k(t_1)+\theta_k(t_2)\right] & \cos\left[\theta_k(t_1)+\theta_k(t_2)\right]
	\end{array}  \right)}.
	\end{eqnarray}
Therefore, if $\theta_k(t)$ is additive in $t$, i.e. $\theta_k(t_1)+\theta_k(t_2) = \theta_k(t_1+t_2)$, then $\lambda_k(t_1)+\lambda_k(t_2) = \lambda_k(t_1+t_2).$ Otherwise, both are simultaneously not additive in $t.$
\end{proof}

\section{Examples}\label{sec:examples}
\subsection{Elementary Physical Processes}
Now we consider some special examples of dynamical maps used in quantum information. They are bit-flipping, phase-flipping, depolarizing, and amplitude damping \cite{NielsenChuang,geobook,NakaharaOhmi}. In the Pauli's basis set $\{\sigma_x,\sigma_y,\sigma_z\},$ the bit-flipping ($BF$) and phase-flipping ($PF$) operations are respectively defined as	
	\begin{eqnarray}
	BF(\rho) &=&  (1-p)\rho + p\sigma_x\rho\sigma_x,\label{eq:BF}\hspace{1em}	PF(\rho) =  (1-p)\rho + p\sigma_z\rho\sigma_z,\label{eq:PF}
	\end{eqnarray}
for $0 < p <1$. They can be written in the real matrix representation as 
	\begin{eqnarray}
	M_{BF} &=&  {\left(\begin{tabular}{ccc} 
	1 & 0 & 0 \\ 
	0 & 1-2p & 0 \\ 
	0 & 0 & 1-2p \\ 
	\end{tabular} \right)},\hspace{1em}	
	M_{PF} =  {\left(\begin{tabular}{ccc} 
	1-2p & 0 & 0 \\ 
	0 & 1-2p & 0 \\ 
	0 & 0 & 1 \\ 
	\end{tabular} \right)}.
	\end{eqnarray}
By their forms, we can see that both are naturally scaling maps. For the bit-flipping, the $x$-axis is the invariant subspace, while the $yz$ plane is the in-plane subspace. For the phase-flipping, the $z$-axis is the invariant subspace, and the $xy$ plane is the in-plane subspace. Both maps do not have a rotation part. In a different manner, the depolarizing operation $(DP)$ is defined by
	\begin{equation}
		DP(\rho) = (1-p)\rho + \frac{p}{2}\iden_2 \label{eq:DP}
	\end{equation}
for $0<p<1,$ so it does not have the splitting of plane and invariant subspaces, which can be seen more evidently from its matrix representation
	\begin{equation}
		M_{DP} =  {\left(\begin{tabular}{ccc} 
	1-p & 0 & 0 \\ 
	0 & 1-p & 0 \\ 
	0 & 0 & 1-p \\ 
	\end{tabular} \right)}.
		\label{eq:MDP}
	\end{equation}
Likewise, the depolarizing process is not composed of any rotation part. The shape of the Bloch ball after the depolarizing process is still a ball but with smaller radius, while those after the bit flipping and phase flipping processes are prolate spheroids with the major axis in the $x$ and $z$ direction, respectively.

Since we do not include a translation in our consideration, some dynamical maps or processes may not be satisfied by the decomposition. A typical example is an amplitude damping process (AD), which is defined by
	\begin{eqnarray}
	\hspace{-2cm} AD(\rho) = & {\left(\begin{tabular}{cc} 
	0 & $\sqrt{p}$  \\ 
	0 & 0  \\
	\end{tabular} \right)} \rho {\left(\begin{tabular}{cc} 
	0 & 0  \\ 
	$\sqrt{p}$ & 0  \\
	\end{tabular} \right)} + {\left(\begin{tabular}{cc} 
	1 & 0  \\ 
	0 & $\sqrt{1-p}$  \\
	\end{tabular} \right)} \rho {\left(\begin{tabular}{cc} 
	1 & 0  \\ 
	0 & $\sqrt{1-p}$  \\
	\end{tabular} \right)}. \label{eq:DA}
	\end{eqnarray}
This mapping is different from the previous examples in that it has no real matrix representation on $\Mat_{d^2-1}(\Real)$ because there is a shift of the center of the Bloch ball by the transformation. This issue may be resolved by mathematical techniques in geometry, such as homogenisation or dimension extension of the real vector space, as mentioned in \ref{appen:SR-TSR}. However, the analysis in this direction is out of scope for this article. Thus, we leave this issue for further investigation. A special case where the space homogenisation technique can be applied together with the unitary-scaling decomposition in the Hilbert-Schmidt representation is demonstrated in the next example.

\subsection{Model of Nuclear Magnetic Resonance}\label{subsec:NMR}
As an illustration, we will consider a specific NMR model in which its Lindblad dynamics reads \cite{NakaharaOhmi}
			{\small\begin{equation}
			\dot{\rho_t}=\Lind(\rho_t) = -i\left[\op{H},\rho_t\right] + \dfrac{1}{2}\sum_{\alpha=+,-,z}\left[\op{L}_\alpha\rho_t \op{L}^\dagger_\alpha-\dfrac{1}{2}\left( \op{L}_\alpha \op{L}^\dagger_\alpha\rho_t +  \rho_t \op{L}_\alpha \op{L}^\dagger_\alpha \right) \right],
			\end{equation}}
\noindent where $\op{H} := -\dfrac{\omega }{2}\sigma_z$ is the background Hamiltonian; $\op{L}_+ :=\dfrac{\sqrt{\Gamma_{+}}}{2}\left(\sigma_x+i\sigma_y \right)$ describes relaxation process; $\op{L}_- := \dfrac{\sqrt{\Gamma_{-}}}{2}\left(\sigma_x-i\sigma_y \right)$ describes excitation process; and $\op{L}_z := \sqrt{\Gamma_z}\sigma_z$ describes dephasing. Here $\Gamma_{+},$ $\Gamma_{-}$ and $\Gamma_{z}$ are relaxation, excitation and dephasing rates, respectively.  Let the initial state be $\rho_0 = \dfrac{1}{2}\left(\iden_2 + x_0\sigma_x + y_0\sigma_y + z_0\sigma_z \right)$ and $T_1^{-1} = \Gamma_{+} + \Gamma_{-}$ and $T_2^{-1} = \dfrac{\Gamma_{+} + \Gamma_{-}}{2} + 2\Gamma_z.$ Then, the evolutionary equation becomes  
			{\footnotesize\begin{equation*}
				\left({\begin{array}{c}
				x_t \\ 
				y_t \\ 
				z_t
				\end{array}}\right) = \left(\begin{array}{ccc}
				e^{-t/T_2}\cos\omega t & -e^{-t/T_2}\sin\omega t & 0 \\ 
				e^{-t/T_2}\sin\omega t & e^{-t/T_2}\cos\omega t & 0 \\ 
				0 & 0 & e^{-t/T_1}
				\end{array}\right) \left({\begin{array}{c}
				x_0 \\ 
				y_0 \\ 
				z_0
				\end{array}}\right) + \left({\begin{array}{c}
				0 \\ 
				0 \\ 
				\dfrac{\Gamma_{+}-\Gamma_{-}}{\Gamma_{+}+\Gamma_{-}}\left( e^{-t/T_1}-1\right)
				\end{array}}\right),
			\end{equation*}}
suggesting that the translation, or the amplitude damping component, cannot be omitted. Since $\Real^3$ can be treated as embedded in the auxiliary space $\Real^4,$ the dynamical matrix of NMR in the $\Real^4$ bases reads
			{\small\begin{equation}
				\M = \left(\begin{array}{cccc} 
				1 & 0 & 0 & 0\\
				0 & e^{-t/T_2}\cos\omega t & -e^{-t/T_2}\sin\omega t & 0 \\ 
				0 & e^{-t/T_2}\sin\omega t & e^{-t/T_2}\cos\omega t &  0 \\ 
				 \dfrac{\Gamma_{+}-\Gamma_{-}}{\Gamma_{+}+\Gamma_{-}}\left( e^{-t/T_1}-1\right) & 0 & 0 & e^{-t/T_1}
				\end{array}\right).
			\end{equation}}
\noindent Again, the role of the unitary-scaling decomposition can be expressed (with indices reshuffled) as
			\begin{equation}
				\M = \left(\begin{array}{cc}
				1 & 0  \\ 
				\dfrac{\Gamma_{+}-\Gamma_{-}}{\Gamma_{+}+\Gamma_{-}}\left( e^{-t/T_1}-1\right) & e^{-t/T_1} 
				\end{array}  \right) \oplus \left( e^{-t/T_2}\rot_{2,z}(\omega t)\right).
			\end{equation}
In particular, when in dynamical equilibrium so that $\Gamma_{+}=\Gamma_{-},$ and the translation part is absent, the block-diagonal is obtained, and the linear entropy yields
			\[\ls(\rho_t) = \dfrac{1}{2} - \left( e^{-t/T_2}(x_0^2+y_0^2) + e^{-t/T_1}z_0^2\right).\]

\end{document}